\documentstyle[twoside,fleqn,espcrc2,epsfig]{article}
\title{\hfill TTP01-01\\
\hfill hep-ph/0101100\\
\hfill January 2001\\
\textbf{
Measuring $\boldmath \sigma(e^+e^-\to {\rm hadrons})$ with Tagged Photons \\
at Electron Positron Colliders\footnotemark
}}
\author{Johann K\"uhn
\address{Institut f\"{u}r Theoretische Teilchenphysik,
Universit\"{a}t Karlsruhe, D--76128 Karlsruhe, Germany}}

\begin{document}
\begin{abstract}
A Monte Carlo generator has been constructed to
simulate the reaction $e^+e^- \to \gamma + 2 \pi$ and $\gamma + 4\pi$
where the photon is assumed to be observed in the detector.
Predictions are presented for cms energies of 1GeV,
3GeV and 10GeV, corresponding to the energies of DA\(\Phi\)NE, BEBC and
of $B$-meson factories.
The event rates are sufficiently high
to allow for a precise measurement of $R(Q^2)$ in the region of $Q$
between approximately 1GeV and 2.5GeV.
Estimates for the kinematic breaking of isospin relations
between different channels 
as a consequence of the $\pi^\pm$ -- $\pi^0$ mass difference
 are given.
\end{abstract}
\maketitle
\newcommand{\Eq}[1]{Eq.(\ref{#1})} 
\newcommand{\labbel}[1]{\label{#1}} 
\newcommand{\be}{\begin{equation}}
\newcommand{\ee}{\end{equation}}
\newcommand{\ba}{\begin{eqnarray}}
\newcommand{\ea}{\end{eqnarray}}

\section{Introduction}

\footnotetext{Talk presented at the Sixth Workshop
on Tau Lepton Physics, Victoria, Canada, 2000. To be published in the
proceedings} 
The precise determination of the cross section for electron positron
annihilation into hadrons over a large energy range is one of the
important tasks of current particle physics. The results are relevant
for the analysis of electroweak precision measurements which are
affected by the running of the electromagnetic coupling from the
Thompson limit up to $M_Z$. Also the prediction for the anomalous
magnetic moment of the muon depends critically on these data. Last not
least the measurement of the energy dependence of R(s) is one of the
gold plated tests of QCD and allows for a precise determination of the
strong coupling constant.

Depending on the energy region different techniques for the measurement
of $R(s)$ have been applied up to date. At low energies, say from the
two pion threshold up to roughly two GeV, exclusive channels are
collected separately. For higher energies inclusive measurements start
to become dominant. For energies below $m_\tau$ isospin invariance and
CVC have traditionally been used to predict $\tau$ decays from electron
positron annihilation \cite{Sakur,Tsai,Gil,KS}. Clearly this
strategy can be inverted \cite{K1}, pending irreducible uncertainties
from isospin violation and radiative corrections \cite{K2}.

To cover a large range of
energies, results from many different experiments and colliders have to
be combined, and energy scans have to be performed to obtain the full
energy dependence. An attractive alternative is provided by the
upcoming $\Phi-$ and $B-$ meson factories which operate at large
luminosities, albeit at fixed energies. Events with radiated tagged
photons give access to a measurement of $R$ over the full range of
energies, from threshold up to the CMS energy of the collider. For
events with tagged photons the invariant mass of the recoiling hadronic
system is fixed by the photon energy which provides an
important kinematic constraint. 

To arrive at reliable predictions including angular and energy cutoffs
as employed by realistic experiments, a Monte Carlo generator 
 is indispensable. For hadronic states with
invariant masses below two or even three GeV it is desirable to simulate
the individual exclusive channels with two, three up to six mesons, 
i.e.\ pions, kaons, etas
 etc.\ which requires a fairly detailed parametrization of the
various form factors.

In principle initial and final state radiation would be required for the
complete simulation. Such a program has been constructed for the two
pion case \cite{BKM}. There it is demonstrated that suitably chosen
configurations, namely those with hard photons at small angles relative
to the beam and well separated from the pions, are dominated by initial
state radiation. In fact, this separation is possible \cite{Graziano} even
when operating the $\phi$ factory DA\(\Phi\)NE on top of the $\phi$ resonance
where direct radiative $\phi$ decays cannot be ignored.  

\begin{table*}
\begin{center}
$$
\begin{array}{||c||c||c||c|c|c||}
\hline
&
&
&
\multicolumn{3}{|c|}{{\rm Event~rates}}\\
\hline
{\rm Collider }& \sqrt{s}
&{\rm Integrated~luminosity}, {\rm fb}^{-1} & \theta_{\rm min} = 5^\circ 
& \theta_{\rm min} = 7^\circ   &  \theta_{\rm min} = 10^\circ 
\\ \hline \hline
{\rm DA}\Phi{\rm NE} & 1.02 & 1 & 13\cdot 10^6 & 12 \cdot 10^6 
&  10 \cdot 10^6 \\ \hline \hline
B-{\rm factory} & 10.6 & 100 & 4 \cdot 10^6 & 3.5 \cdot 10^6 
&  3 \cdot 10^6 \\ \hline \hline
{B-\rm factory} & 10.6 & 100 & 2.7\cdot 10^6 & 2.3\cdot 10^6
&  2.0\cdot 10^6 \\ 
\hline
\end{array}
$$
\caption{Estimated number of  radiative 
events $e^+e^- \to \ hadrons \ + \ \gamma$ for different center of mass
energies. In the first two rows \(hadrons\) stands for
\(\pi^+\pi^-\) and 
the minimal photon energy is $0.1$ GeV. The third row is 
 obtained for a continuum
 contribution in the region
  \(2 \  {\rm GeV} < \sqrt{Q^2} < 3.7 \ {\rm GeV}\)
 assuming a constant \(R=2.4\).  }
\end{center}
\label{t1}
\end{table*}
In a recent paper \cite{CzKu} this project has been continued
with the construction of a
generator for the radiative production of the four pion final state,
including the $\omega (\to3\pi) \pi$ channel. This mode contributes a
large fraction of the rate with invariant masses
  between one and two GeV. The
energy region between 1.5 GeV and 2.5 GeV is difficult to access 
directly with
current electron positron colliders. At the same time the experimental
uncertainties are relatively large. This motivates the special effort
devoted to this range.

The Monte Carlo program 
is constructed in a modular form such that the
parametrization of the hadronic matrix element can easily be replaced by
a more elaborate version. Different final states with three, four
or five pions or kaons can be included. The present
parametrization of the hadronic matrix element follows closely the form
suggested in \cite{Fink}, correcting only some minor deficiencies.
 The four pion amplitude is assumed to be dominated by
 \(\rho\prime \to \pi a_1 \) plus a direct coupling 
 \(\rho\prime \to \rho\pi\pi\) and exhibits the proper behavior in the chiral
 limit.

\section{The radiative return}

  Hard photons observed at small angles relative to the electron or
positron beam and at the same time well separated from charged particles
in the final state can be used to reduce the effective center of mass
energy at electron positron colliders. 
As shown in Table \ref{t1}, the event rates are fairly high, of 
${\cal O}(10^6)$ both at DAPHNE and at B-meson factories. 
Performing a detailed analysis of
the angular and energy distributions for the $\gamma \pi^+ \pi^-$ final
state it has been shown that initial and final state radiation can be
reasonably well separated \cite{BKM,Graziano}. 
For the four pion case we
therefore restrict the discussion to initial state radiation only. The
amplitude for the production of an arbitrary hadronic final state
\begin{eqnarray}
\lefteqn{
 {\cal M} = i \ e^3 \ \bar v\left(p_{+}\right) 
 \left[ \gamma^{\nu} \frac{1}{p_{-}{\kern-12pt}/ - k {\kern-5pt}/ - m}
  \epsilon^{*}{\kern-10pt}/\left(k \right)\right.
}\\
 && \left.+ \epsilon^{*}{\kern-10pt}/\left(k \right) 
    \frac{1}{k {\kern-5pt}/ - p_{+}{\kern-12pt}/ - m}\gamma^{\nu}
 \right] \frac{1}{Q^2}\ J^{em}_{\nu} \nonumber
\end{eqnarray}
involves the matrix element of the hadronic current
\begin{eqnarray}
J^{em}_{\nu} & \equiv & J^{em}_{\nu}\left(q_1,...,q_n\right)\nonumber\\ 
             & \equiv & <h(q_1),...,h(q_n)|J^{em}_{\nu}\left(0\right)|0>
\end{eqnarray}
which has to be parameterized by form factors to be discussed below. For the
two pion case the amplitude is determined by only one function, the pion
form factor \(F_{2\pi}\). 
\begin{eqnarray}
J_{\nu}^{em,2\pi} 
 = \left(q_{\nu}^{+}-q_{\nu}^{-}\right)F_{2\pi}\left(Q^2\right)
\end{eqnarray}
The matrix element for the four pion case will be discussed below.
After integrating
the hadronic tensor $H_{\mu\nu}$ over the hadronic phase space one gets
\begin{eqnarray}
\lefteqn{
 \int \ J^{em}_\mu (J^{em}_\nu)^*  \ \ d\bar\Phi_n(Q;q_1,\dots,q_n) =}
\\ 
 &&\hfill
\frac{1}{6\pi} \left(Q_{\mu}Q_{\nu}-g_{\mu\nu}Q^2\right) \ R(Q^2)\nonumber
\labbel{rr}
\end{eqnarray}
where \(R(Q^2)\) is \(\sigma(e^+e^-\rightarrow hadrons)/\sigma_{point}\).
The additional integration of the differential cross section 
over the photon angles
(the azimuthal angle is integrated over the full range and
the polar angle within \(\theta_{min} < \theta < \pi-\theta_{min}\) )
leads to the differential distribution 
\begin{eqnarray}
\lefteqn{
Q^2 \frac {{\rm d}\sigma}{{\rm d} Q^2} = \frac{4 \alpha^3}{3 s} R(Q^2)
\left \{ \frac {(s^2+Q^4)}{s(s-Q^2)} 
\log \frac {1+\cos \theta_{\rm min}}{1-\cos \theta_{\rm min}}\right.}
\nonumber\\
& &\left.-\frac {(s-Q^2)}{s}\cos \theta_{\rm min} \right \} \ ,
\labbel{1}
\end{eqnarray}
which can be used to calculate the event rate observed for realistic photon
energy and angular cuts (see Tab.1).

\section{Isospin relations}

The emphasis of \cite{CzKu} was towards final states consisting of
 four pions and a photon. Before entering a discussion of a model 
 dependent parametrization of the form factors 
 the constraints from isospin invariance must be recalled. 
They relate the amplitudes
 of the \(e^+ e^- \to 2\pi^+2\pi^-\) and \(e^+ e^- \to \pi^+\pi^- 2\pi^0\)
 processes and those for \(\tau\) decays into \(\pi^-3\pi^0\) and
 \(\pi^+2\pi^-\pi^0\). The amplitude for  \(\tau\) decay 
\be
 {\cal M}_{\tau} =  \frac{G_F}{\sqrt{2}} \ \cos\theta_c \ \
 \bar v\left(p_{\nu}\right) \gamma^{\alpha}\left(1-\gamma_5\right)
 u\left(p_{\tau}\right) \ \ J_{\alpha}^{-} 
\ee
leads to the differential distribution
\begin{eqnarray}
\lefteqn{\frac{d\Gamma}{dQ^2} =}\\
&& 2 \ \Gamma_e \frac{\cos^2\theta_c}{m_{\tau}^2}
 \left(1-\frac{Q^2}{m_{\tau}^2}\right)^2 
 \left(1+2\frac{Q^2}{m_{\tau}^2}\right) R^{\tau}\left(Q^2\right) \nonumber
\labbel{diff}
\end{eqnarray}
with 

\begin{eqnarray}
\lefteqn{
 \int \ J_\mu^- J^{-*}_\nu  \ \ d\bar\Phi_n(Q;q_1,\dots,q_n) =} \\ 
&& \frac{1}{3\pi} \left(Q_{\mu}Q_{\nu}-g_{\mu\nu}Q^2\right) \
 R^{\tau}(Q^2) 
\nonumber
\labbel{rrt}
\end{eqnarray}
Note the relative factor of 2 between the definitions in \Eq{rr} and \Eq{rrt}.

Final states with an even number of pions are produced through the isospin
one part of the electromagnetic current only, whence

\begin{eqnarray}
 \sqrt{2} \ \ J^{em}_{\mu}(2\pi) = J^{-}_{\mu}(2\pi)
 \labbel{rel0}
\end{eqnarray}

\noindent
and \(R(Q^2)=R^{\tau}(Q^2)\) for two pion final states.
 
A similar relation for the four pion final state is easily obtained
\cite{K2,CzKu}:
\ba
\lefteqn{\langle \pi^+ \pi^- \pi_1^0 \pi_2^0 | J^3_{\mu} | 0 \rangle = 
J_{\mu}(p_1,p_2,p^+,p^-)
} \nonumber \\
\lefteqn{
\langle \pi^+_1 \pi^+_2 \pi^-_1 \pi^-_2 | J^3_{\mu} | 0 \rangle =
}\nonumber\\
&& J_{\mu}(p_2^+,p_2^-,p_1^+,p_1^-) + J_{\mu}(p_1^+,p_2^-,p_2^+,p_1^-) 
\nonumber \\
&+&J_{\mu}(p_2^+,p_1^-,p_1^+,p_2^-)+ J_{\mu}(p_1^+,p_1^-,p_2^+,p_2^-)
\nonumber \\
\lefteqn{
\langle \pi^- \pi^0_1 \pi^0_2 \pi^0_3 | J^{-}_{\mu} | 0 \rangle =
J_{\mu}(p_2,p_3,p^-,p_1)+
}\nonumber\\
&&J_{\mu}(p_1,p_3,p^-,p_2)+J_{\mu}(p_1,p_2,p^-,p_3)
\nonumber \\
\lefteqn{
\langle \pi^-_1 \pi^-_2 \pi^+ \pi^0 | J^{-}_{\mu} | 0 \rangle =
}\nonumber\\ 
&&J_{\mu}(p^+,p_2,p_1,p^0)+ J_{\mu}(p^+,p_1,p_2,p^0)
 \labbel{rel} \ ,
\ea
\noindent
which connects \(\tau\) decay and electron positron annihilation
($J^{em}=\frac{1}{\sqrt{2}}J^3$).
The (in $Q^2$) differential rates are then
consistent with the familiar relations between \(\tau\) decays and
 \(e^+e^-\) annihilation into four pions:
\ba
R^{\tau}\left(- 0  0  0\right)& =&  \frac{1}{2}  R\left(+ +  -  -\right)\\
R^{\tau}\left(- -  +  0\right) 
& =& \frac{1}{2}  R\left(+ +  -  -\right) +  R\left(+ -  0  0\right)\nonumber
\labbel{CVC}
\ea
\begin{table*}
\begin{center}
$$
\begin{array}{|c|c|c|}
\hline
  
& {\rm only} \  \beta^3  \ {\rm modified }
&   \beta^3 \  {\rm and}  \ \Gamma_{\rho} \ {\rm modified }\ \\
\hline
 0.28\  < \ E \ [GeV] \ < 0.81 &       &       \\
 a_{\mu}                       &1.0163 &1.0088 \\
          \Delta \alpha(M_Z)   &1.0116 &1.0027 \\
\hline
0.32\  < \ E \ [GeV] \ < 1.777 &       &  \\
a_{\mu}                        &1.0130 &1.0058 \\
\Delta \alpha(M_Z)             &1.0096 &1.0016 \\
\hline
\end{array}
$$
\caption{
Kinematic correction factors for the predictions of \(a_{\mu}\)
 and \(\Delta\alpha\) from \(\tau\) data. 
 }
\end{center}
\label{tspec}
\end{table*}

\section{The \(\pi^{\pm}\)-\(\pi^0\) mass difference}
 All the relations obtained in the previous section are strictly
 applicable only
 in case all pions in the final states have the same mass, which is
 obviously not true. The relatively large (about 3.6\%) 
\(\pi^{\pm}\)-\(\pi^0\) mass difference will affect the 
 \(R(Q^2)\leftrightarrow R^\tau(Q^2)\) relations even if 
 \Eq{rel0} and \Eq{rel} still hold. 
Let us retain these relations between the amplitudes and 
incorporate the \(\pi^{\pm}\)-\(\pi^0\) mass difference in the phase
space considerations. This should give at least 
 an indication of the size of these "kinematic" isospin violations.

To estimate these effects for the two-pion case
assume that the cross section \(\sigma(e^+e^-\to\pi^+\pi^-)\)
 is well measured and the squared form factor 
 \ba
   |F(Q^2)|^2 = \frac{3 Q^2}{\pi \alpha^2 \beta_{\pi}^3} 
 \ \sigma(e^+e^-\to\pi^+\pi^-)(Q^2) \ ,
 \ea
\( \left( \beta_{\pi} 
 = \left(1 - 4 m_{\pi^-}^2 / Q^2 \right)^{1/2}  \right)\)
then used to predict the \(\tau^-\to\nu_{\tau}\pi^-\pi^0\) decay rate
The size of the corrections depends critically on the details of 
the assumptions.

If the form factor and the form of the current 
\ba
 J^{em}_\mu(2\pi) = \sqrt{2} \ J^-_\mu(2\pi) = (q_{1,\mu}-q_{2,\mu})
 \ F(Q^2) 
 \labbel{cur1}
\ea 
remain unchanged, the integral for the \(\tau\) rate receives an
S-wave contribution which can be eliminated by replacing \Eq{cur1} by
\ba
\lefteqn{ 
\sqrt{2} \ J^-_\mu(2\pi) = 
} \\
\lefteqn{
\left(q_{1,\mu}-q_{2,\mu} 
   - \frac{q_{1,\mu}+q_{2,\mu}}{Q^2}\ Q.(q_1-q_2)\right) \ F(Q^2)} \nonumber
\ea 
Numerically the contribution to the integral of this latter term is tiny
 -- nevertheless we shall adopt the second choice.

The introduction of the $\pi^\pm-\pi^0$ mass difference
raises the prediction for the \(\tau\) decay rate
 by 0.86\%. It seems, however, plausible, that the energy dependent width
 of the \(\rho\)- meson, which is present in the form factor, has to be modified
 accordingly, 
 leading to an effective increase of \(\Gamma_{\rho}\) by 0.74\%.
The two effects nearly compensate in the
 integral. Hence the relation between \(\tau^- \to \nu_{\tau} \pi^- \pi^0 \) 
 partial decay width and the \(e^+e^-\to\pi^+\pi^-\) cross section
 would only be corrected by 0.06\%.
 However, a sizable \(Q^2\) dependence
 of the ratio of the two spectral functions
 (\(V(Q^2) \sim |F(Q^2)|^2 \beta^3\) )
 is expected, with a reduction approximately 0.74 \% close to the peak
 of the \(\rho\) resonance and enhancements at the tails. 

 At present, however, \(\tau\) data provide an important input for the 
 prediction of the QED coupling at the scale of \(M_Z\) and the hadronic
 contribution to g-2 \cite{K1,ADH}. Let us, for the moment, assume that
 the aforementioned kinematic effects are indeed present. If the 
 \(e^+e^-\) cross section is deduced from \(\tau\) data through \Eq{CVCc}
\begin{eqnarray}
\lefteqn{
\frac{1}{\Gamma_e} 
\frac{d\Gamma(\tau^-\to \nu_\tau\pi^-\pi^0)}{dQ^2} =
\frac{3\cos^2\theta_c} {2\pi \alpha^2 m_{\tau}^2}
}\\
\lefteqn{
 Q^2 \ \left(1-\frac{Q^2}{m_{\tau}^2}\right)^2 
 \left(1+2\frac{Q^2}{m_{\tau}^2}\right) 
 \ \sigma(e^+e^-\to\pi^+\pi^-)
}\nonumber 
\labbel{CVCc}
\end{eqnarray}
and the contributions to g-2 and \(\alpha(M_Z)\) are evaluated without
 kinematic corrections of phase space and form factor 
 the former
 are overestimated by 0.58\% and 0.16\%
 respectively (Tab.2).

The situation is more complicated for the four pion case. 
To estimate the size of the effect the quantities \(R(--+0)\) etc. 
were evaluated once assuming that all masses are equal to \(m_-\) 
and once taking the real masses. 
The corrections to the integrated quantities 
modify the integrated version of the \Eq{diff} accordingly:
 \ba
\frac{1}{1.050} \Gamma(-000)& =& \frac{1}{2}  \Gamma(++--)
 \\
 \frac{1}{1.024} \Gamma(--+0)& =& \frac{1}{2}  \Gamma(++--)
\\
&& + \frac{1}{1.046} \Gamma(+-00)
\nonumber
\ea
\section{The hadronic current}

\begin{figure*}
\begin{center}
\epsfig{file=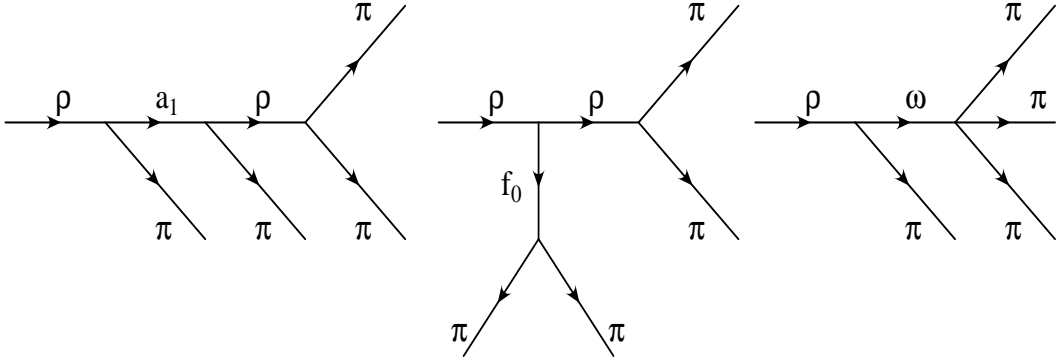,width=14cm,height=5cm}
\end{center}
\caption{Diagrams contributing to the hadronic current.
}
\label{f2}
\end{figure*}

Employing isospin relations it is thus sufficient to construct
the hadronic current for the \((+-00)\) mode only. The basic building block
of the current contains a part built on the assumption of  \( a_1\) vector
dominance plus an \(\omega\) exchange contribution. By adding an \(f_0\)
contribution one indeed recovers the proper chiral limit \cite{FWW}.
The three contributions are depicted schematically in Fig.\ref{f2} 
and described in detail in \cite{CzKu}.
\begin{table*}
\begin{center}
$$
\begin{array}{|c|c|c|c|}
\hline
 {\rm Mode} 
& {\rm \cite{Fink}}
& {\rm present \  model}
& {\rm experiment}\\
\hline
 {\rm Br}\left( \tau^-\to\nu_{\tau}2\pi^-\pi^+\pi^0 \right)
&3.11\%
&4.33\%
&4.20(8)\% \\
\hline
 {\rm Br}\left( \tau^-\to\nu_{\tau}\pi^-
  \omega\left(\pi^-\pi^+\pi^0\right) \right) 
&1.20\%
&1.48\%
&1.73(11)\% \\
\hline
 {\rm Br}\left( \tau^-\to\nu_{\tau}\pi^-3\pi^0 \right)
&0.98\%
&1.14\%
&1.08(10)\% \\
\hline
\end{array}
$$
\vspace*{0.5cm}
\caption{
 Branching ratios of \(\tau\) decay modes. Results of \cite{Fink} 
 and the present current are compared to experimental data. 
 }
\end{center}
\label{t3}
\end{table*}

To test the physical predictions of the current, let us start
 with the \(\tau\) decay branching ratios,
 summarized in Tab.3. The agreement between the predictions of the current
 \cite{CzKu}  and the data is satisfactory for the 
 \(\tau^-\to\nu_{\tau}\pi^-3\pi^0\) decay mode.
Comparing however the results for the  
 \(\tau^-\to\nu_{\tau}2\pi^-\pi^+\pi^0\) and
  \(\tau^-\to\nu_{\tau}\pi^-
  \omega\left(\pi^-\pi^+\pi^0\right)\)
  modes
 it seems that the \(\omega\)
 part of the current does not represent the data well, even if
 the total branching ratio for the \(\tau^-\to\nu_{\tau}2\pi^-\pi^+\pi^0\)
 decay mode agrees with the data.

In \cite{CzKu} also the predictions for the \(e^+e^-\to 2\pi^+ 2\pi^-\)
and \(e^+e^-\to \pi^+\pi^-2\pi^0\) cross sections are compared with data,
and satisfactory agreement between the Monte Carlo and the data is observed.

\section{The Monte Carlo program}
 The idea behind the structure of the Monte Carlo program
 is to allow for a simple addition of new final state modes into the program
 and for a simple replacement of the current(s) of the existing modes.
 The program thus exhibits a modular structure. For
 the generation of the four momenta of the mesons no sophisticated method
 of a variance reduction was applied. This slows down the generation,
 but could be accounted for if a faster Monte Carlo generator 
 would be required.
 It has, however, the advantage of being universal and no
 change of the variance reduction method is required
 with each modification of the hadronic
 current.
 The process to be simulated by the program in its final stage is
 \(e^+ e^- \rightarrow \gamma \ + \ hadrons \) 
 with an exclusive description of final states,
even if till now
 only  \(\pi^+ \pi^-\), \(2\pi^0 \pi^+ \pi^-\) and \(2\pi^+ 2\pi^-\)
 hadronic final states are implemented. The LL radiative QED
 corrections were taken into account using structure function method
 as developed in \cite{CCR} and limited to the initial emission only.
 In fact the program can run in one of two modes (chosen by a user)
  one with collinear radiation
 and one without it.
 Hard large angle photon 
 emission is limited to initial state radiation,
  which is justified by \cite{BKM} where it was demonstrated for the 
 \(\pi^+ \pi^-\) hadronic state that the contribution from the final
 state emission as well as the initial-final state interference can be
 reduced to a negligible level by applying suitable cuts.
 
 Let us now discuss the cuts, which
 reduce the contribution of final state radiation to a negligible level.
 We recall that in \cite{BKM} it was shown that the following set of 
 angular cuts 
 
\ba
\lefteqn{\rm cuts1:}\\
&&
 \left( 7^{\circ} < \theta_\gamma < 20^{\circ} 
 \ \ \ \ \ or \ \ \ \ \ 160^{\circ} < \theta_\gamma < 173^{\circ}\right) 
\nonumber\\
&&
 \ \ \  and 
  \ \ \ 30^{\circ} < \theta_\pi < 150^{\circ} \ ,\nonumber
 \labbel{cuts}
\ea

\noindent
 ( \(\theta_\gamma\) (\(\theta_\pi\) ) is the photon (pion) polar angle)
 fulfils this requirement for the 
 \(e^+ e^- \to \pi^+ \pi^- \gamma\) cross section at DAPHNE. 
It reduces, however,
 the observed cross section significantly. This starts to become dramatic, when
 one runs at energies well above 1 GeV. The following
 set of cuts

\ba
 \lefteqn{\rm cuts2:}\\
 &&\left( 7^{\circ} < \theta_\gamma < 20^{\circ}   \ \
 and  \ \  30^{\circ} < \theta_\pi < 173^{\circ}\right) \nonumber \\
   &&or  \ 
 \left( 160^{\circ} < \theta_\gamma < 173^{\circ}  \ 
 and  \  7^{\circ} < \theta_\pi < 150^{\circ}\right) \nonumber
\labbel{cuts1}
\ea

\noindent
also reduces the contribution from final state radiation
 to a
 negligible level due to the fact that
 the pions and photon are well separated as in the
 previous case.

\begin{figure}[htbp]
\epsfig{figure=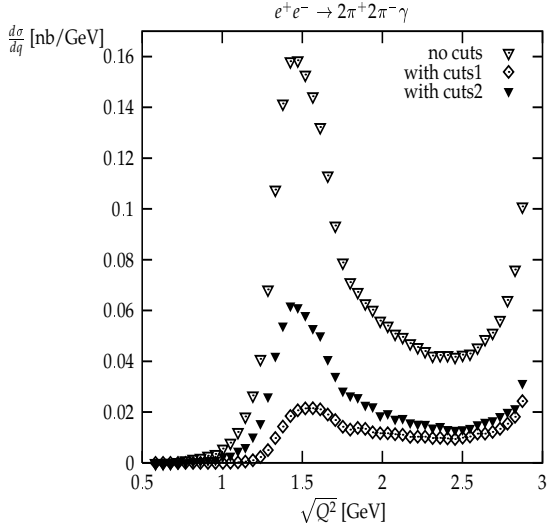, width=0.45\textwidth,height=7cm}
\caption{The differential \(e^+e^-\to 2\pi^+2\pi^-\gamma\) 
cross sections for a
 beam energy of 1.5 GeV and a minimal photon energy of 0.1 GeV for different
cuts as described in the text.
}
\label{f5}
\end{figure}

At the same time the cross section reduction is much
 smaller, especially for higher beam energies and higher energies
 of the observed photons. The effect of the two sets of cuts on the
 cross sections with four charged pions in the final state is presented 
in Fig.\ref{f5} for  \(2E_{beam}\) = 3 GeV. ``No cuts'' corresponds to
 $7^\circ \le \theta_\gamma\le 173^\circ$ and not cuts on the pions.  
 For higher beam energies the effect of the cross section
 reduction is much stronger and at 2 \(E_{beam}\) = 10 GeV the cross
 sections for the cuts specified in \Eq{cuts} is reduced almost to zero. 
 For the cuts specified in \Eq{cuts1} the  reduction remains tolerable
 (Table \ref{t4}).

\begin{table*}
\begin{center}
$$
\begin{array}{||c||c||c|c||}
\hline
&
&
\multicolumn{2}{|c|}{{\rm Event~rates}}\\
\hline
\sqrt{s}
&{\rm Integrated~luminosity}, {\rm fb}^{-1} 
& \ \ \ 2\pi^+2\pi^-\gamma \ \ \ 
& \ \ \ 2\pi^0\pi^+\pi^-\gamma \ \ \ 
\\ \hline \hline
\ 1 \ {\rm GeV}\ & 1 & 1.04\cdot 10^4 & 1.13 \cdot 10^4 
 \\ \hline \hline
\  3\ {\rm GeV}\ & 1 & 4.66 \cdot 10^4 & 5.72 \cdot 10^4 
  \\ \hline \hline
\  10\ {\rm GeV}\  & 100 & 1.86 \cdot 10^5 & 2.33 \cdot 10^5 \\ 
\hline
\end{array}
$$
\caption{Estimated number of  radiative 
events $e^+e^- \to \ 4\pi \ + \ \gamma$ for different center of mass
energies. The minimal photon energy is: 0.05 GeV (first row),
 0.1 GeV (second row), 0.2 GeV (third row). The angular cuts of \Eq{cuts1}
 were applied.
 }
\end{center}
\label{t4}
\end{table*}
One may even restrict photon and pions detection angles to the central
 region, e.g. 
 \( 25^{\circ} < \theta_\gamma < 155^{\circ}\)
 and   \( 30^{\circ} < \theta_\pi < 150^{\circ}\) respectively
 if a minimal angle of \(20^{\circ}\) between photon and charged 
 and neutral pions is required in order to suppress final state
 radiation and to clearly separate neutral pions and the photon.
 With this cut one obtains (\(\sqrt{s}= 10 \ GeV\) and \({\cal L}
 = 100 \ fb^{-1} \) ) a rate of \(1.17\cdot 10^5\) events with
 \(2\pi^+ 2\pi^- \gamma\) and \(1.53 \cdot 10^5\) events with
 \(2\pi^0\pi^{+}\pi^{-}\gamma\).

 Additional collinear emission always present in the real experiment
 reduces slightly the cross sections.
 Its actual size depends on the cuts on the invariant mass of
 the \(4\pi\ +\ \gamma\) system. The effect is similar for different
 energies and for both charge modes.

\section{Summary}
As an alternative to a direct measurement of the cross section for 
$e^+e^-\to {\rm hadrons}$ at the relevant energy one may use initial state
radiation to reduce the effective energy of electron positron colliders, 
exploiting the large luminosity of ``factories'' and accessing thus a
continuum of hadronic final states.

With this motivation a Monte Carlo generator has been constructed to
simulate the reaction $e^+e^- \to \gamma + 2 \pi$ and $\gamma + 4 \pi$, 
where the photon is assumed to be observed in the detector. 
Once more accurate data become available, the modular
structure of the program will allow for modification or
replacement of the hadronic current in a simple way. Additional
collinear photon radiation has been incorporated with the technique of
structure functions. 

Predictions are presented for cms energies of 1GeV,
3GeV and 10GeV, corresponding to the energies of DA\(\Phi\)NE, BEBC and
of $B$-meson factories.
Even after applying realistic cuts the event rates are sufficiently high
to allow for a precise measurement of $R(Q^2)$ in the region of $Q$
between approximately 1GeV and 2.5GeV.


\vskip 0.4 cm

{\bf Acknowledgments.}
The author would like to thank H. Czy\.z for the pleasant collaboration
on the topics discussed in this talk.
Work supported by BMBF under grant BMBF-057KA92P. 

\end{document}